# OVERVIEW OF STREAMING-DATA ALGORITHMS

T. Soni Madhulatha

Department of Informatics, Alluri Institute of Management Sciences, Warangal, A.P.
latha.gannarapu@gmail.com

*ABSTRACT*

*Due to recent advances in data collection techniques, massive amounts of data are being collected at an extremely fast pace. Also, these data are potentially unbounded. Boundless streams of data collected from sensors, equipments, and other data sources are referred to as data streams. Various data mining tasks can be performed on data streams in search of interesting patterns. This paper studies a particular data mining task, clustering, which can be used as the first step in many knowledge discovery processes. By grouping data streams into homogeneous clusters, data miners can learn about data characteristics which can then be developed into classification models for new data or predictive models for unknown events. Recent research addresses the problem of data-stream mining to deal with applications that require processing huge amounts of data such as sensor data analysis and financial applications. For such analysis, single-pass algorithms that consume a small amount of memory are critical.*

## 1. INTRODUCTION

For many recent applications, the concept of a data stream is more appropriate than a data set. By na-ture, a stored data set is an appropriate model when significant portions of the data are queried again and again, and updates are small and/or relatively infrequent. In contrast, a data stream is an appropriate model when a large volume of data is arriving continuously and it is either unnecessary or impractical to store the data in some form of memory. Some applications naturally generate data streams as opposed to simple data sets.

**Motivation**

Time series data mining involves the discovery of interesting patterns from time series data that were previously unknown to information users. Through the use of time series data mining, researchers are able to perform various tasks on time series data, such as time series classification, time series clustering, rule extraction and pattern querying. To perform these tasks, different techniques have been established.
With substantial growth in computer network during the past few decades, time series data are being collected continuously. Time series data of such nature are called data streams, which flow through computer systems at high speeds. Data streams have greater impact on computing systems than static time series data. Challenges introduced by time series data are faced on a greater scale, due to larger data volumes. New challenges are also presented by data streams. The data streams often are too large to fit in the main memory, thus greatly affecting data streams processing systems on how streams are processed. There is a real need for effective data streams mining techniques that can handle the challenges of data streams efficiently.

## 1.2. Time series data

Time series data constitutes a large portion of the data stored in real world databases. Time series data appears in many application domains: financial, weather, medical, social science, computer networks, and business. This time series data is derived from recording the observations of various types of measurements, e.g., temperature, stock price, household income, patient's heart rate, number of bits transferred, and product's sale volume over a period of time. Some complex data types, such as audio and video, are also considered time series data, since they can be measured at each point in time.

## 1.3. Data Streams Clustering

Mining time series data and data streams have been recognized as two of the top ten research problems in data mining. Among all data mining techniques, clustering can be used as the first step in the data mining process. By grouping data streams into homogeneous clusters that are described based on similarities, data miners can learn about data characteristics which can then be developed into classification models for new data or predictive models for unknown events.
Clustering algorithms in general have been categorized into five types: partitioning, hierarchical, density-based, grid-based, and model-based. Clustering algorithms that are frequently cited in the literature include K-means, CLARANS, BIRCH, DBSCAN, and STING. However, challenges arise when these clustering techniques are applied to time series data. Because time series data has its unique characteristics of having very high dimensionality (number of time points), executing clustering algorithms become costly in terms of computational time. Various approaches exist in clustering time series with high-dimensionality. Related work on clustering time series data relies on transforming raw time series data using dimensionality reduction techniques, and then performing clustering on the transformed data to resolve the dimensionality problem. However, in data streams clustering the need of online techniques that can cluster data points incrementally arises.

In computer science, data stream clustering is defined as the clustering of data that arrive continuously such as telephone records, multimedia data, financial transactions etc. Data stream clustering is usually studied under the data stream model of computation and the objective is, given a sequence of points, to construct a good clustering of the stream, using a small amount of memory and time. Data stream clustering has recently attracted attention for emerging applications that involve large amounts of streaming data.

## 2. Definitions

The existing approaches of data streams clustering in literature and various related issues. First, it examines the challenges of dealing with the static time series data type and how these challenges are overcome in the context of clustering. It also investigates the challenges of managing transient data streams for data mining, followed by the issues that arise when clustering data streams are considered. As a result, the requirements of an eligible data streams clustering technique are reviewed.

### 2.1. Definitions of Time Series Data and a Time Series Dataset

A time series is a sequence of event values which occur during a period of time. Each event occurring at each time point has a value which is recorded. The collection of all these values represents a single variable such as EEG signal, or stock price over a time period. Therefore, a time series of a single variable contains a sequence of recorded observations of an interesting event. Formally a time series can be represented by:

$$S = \{s_1, s_2, \ldots, s_n\} \text{------------------------------------------(1)}$$

where $S$ is a whole time series, $s_i$ is the recorded value of variable $s$ at time $i$, and $n$ is the number of observations. A time series dataset $D$ consists of a collection of time series $S_j$, when $1 \leq j \leq m$, and $m$ is the number of time series.

Time series data appears in many application domains, such as in financial, meteorological, medical, social sciences, computer networks, and business. Time series are derived from recording the observations of various types of phenomena, e.g., temperature, stock prices, household income, patient's heart rate, number of bits transferred, and product's sale volume over a period of time. Some complex data types, such as audio and video, are also considered time series data, since they can be measured at each point in time.

## 2.2. Definition of Data Streams Clustering

In a data streams clustering context, there are a fixed $m$ number of data sources such as geographical sensors that are sending temperature data at given periodic intervals. We assume that $n$ number of data observations from each data source will be transmitted from time $T_1$ to $T_n$. In addition to a static time series dataset $D$, $n$ can increase indefinitely in a data streams environment. A supporting data streams clustering system will continuously handle all data points flowing into the system from $m$ data sources for $n$ time intervals, one interval at a time.

## 3. RELATED ALGORITHMS

Partitioning methods subdivide a dataset into k groups. One such example is the k-medoids algorithm which selects k initial centers, repeatedly chooses a data point randomly, and replaces it with an existing center if there is an improvement in SSQ. k-medoids is related to the CG algorithm that solves the facility location variant which is more desirable since in practice one does not know the exact number of clusters k. Choosing a new medoid among all the remaining points is time-consuming; to address this problem, CLARA used sampling to reduce the number of feasible centers. A distinguishing feature of our approach is a careful understanding of how sample size affects clustering quality. CLARANS draws a fresh sample of feasible centers before each calculation of SSQ improvement. All of the k-medoid types of approaches, including PAM, CLARA, and CLARANS, are known not to be scalable and thus are not appropriate in a streaming context. For clustering, k-means is a widely used heuristic but alternate algorithms have also been developed such as k-medoids, CURE and the popular BIRCH.

### K-medoid

The basic strategy of k-medoids algorithm is each cluster is represented by one of the objects located near the center of the cluster. PAM (Partitioning around Medoids) was one of the first k-medoids algorithm is introduced.

*k*-medoid clustering algorithm:
1. Initialize: randomly select *k* of the *n* data points as the medoids
2. Associate each data point to the closest medoid.
   closest here is defined using any valid distance metric, most commonly Euclidean distance, Manhattan distance or Minkowski distance.
3. For each medoid *m*
    1. For each non-medoid data point *o*

    2. Swap *m* and *o* and compute the total cost of the configuration
4. Select the configuration with the lowest cost.
repeat steps 2 to 5 until there is no change in the medoid

## K-Means

The k-means method is the standard clustering algorithm, described by Hartigan and still enjoys widespread use. The method partitions the data into k clusters, where the k is supplied by the user. Clusters are described by the p-vector mean of the objects contained in the cluster, referred to as a centroid.

The algorithm partitions the objects so as to minimize the within-cluster discrepancies, where a discrepancy is defined as the difference between an object and its centroid. The k-means algorithm is as follows:
1. Create k centroids to initialize the algorithm.
2. Assign each of the n objects to the cluster for which it has the smallest L2 norm.
3. Update the centroids in light of the current membership.
4. For each object i, where $x_i \in C_j$ calculate:

$$h = \arg\min_{r \neq j} \frac{n_r \|x_i - c_r\|}{n_r - 1} \quad \ldots\ldots\ldots\ldots\ldots\ldots\ldots\ldots\ldots \quad (2)$$

where $n_r$ is the number of objects assigned to cluster r.

If $\dfrac{n_h \|x_i - c_h\|}{n_h - 1} < \dfrac{n_j \|x_i - c_j\|}{n_j - 1}$ then move object i from cluster j to cluster h.

Update the values of the relevant centroids.

    4. If an object has been moved in the last n calls to Step 4, go to Step 3. Otherwise stop.

## 3.3 CLARA

An obvious way of clustering larger datasets is to try and extend existing methods so that they can cope with a larger number of objects. The focus is on clustering large numbers of objects rather than a small number of objects in high dimensions. Kaufman and Rousseau (1990) suggested the CLARA: Clustering for Large Applications algorithm for tackling large applications. CLARA extends their k-medoids approach for a large number of objects. It works by clustering a sample from the dataset and then assigns all objects in the dataset to these clusters.

This algorithm relies on the sampling approach to handle large data sets. Instead of finding medoids for the entire data set, CLARA draws a small sample from the data set and applies the PAM algorithm to generate an optimal set of medoids for the sample. The quality of resulting medoids is measured by the average dissimilarity between every object in the entire data set D and the medoid of its cluster, defined as the following cost function:

$$Cost(M,D) = \frac{\sum_{i=1}^{n} dissimilarity(O_i, rep(M, O_i))}{n} \quad \ldots\ldots\ldots\ldots\ldots\ldots\ldots\ldots..(3).$$

where M is a set of selected medoids, dissimilarity($O_i$, $O_j$) is the dissimilarity between objects $O_i$ and $O_j$, and rep(M, $O_i$) returns a medoid in M which is closest to $O_i$.

**The algorithm is as follows:**

1. Draw a sample from the n objects and cluster it into k groups.
2. Assign each object in the dataset to the nearest group.
3. Store the average distance between the objects and their respective groups.
4. Repeat the process five times, selecting the clustering with the smallest average distance.

While providing a means to assign a large number of objects to groups, CLARA is clearly not ideal. Let M be the maximum number of objects that our clustering method can process in a reasonable time. In cases where n >> M, clustering a small sample from the data will often result in some groups present in the data being missed entirely.

## 3.4 CLARANS

Instead of exhaustively searching a random subset of objects, CLARANS proceeds by searching a random subset of the neighbours of a particular solution, S. Thus the search for the best representation is not con_ned to a local area of the data. The CLARANS algorithm is governed by two parameters: $MAX_{neigh}$, the maximum number of neighbours of S to assess; and $MAX_{sol}$, the number of local solutions to obtain.

The CLARANS algorithm is as follows:
1. Set S to be an arbitrary set of k representative objects. Set i = 1.
2. Set j = 1.
3. Consider a neighbour R of S at random. Calculate the total swap contribution of the two neighbours.
4. If R has a lower cost, set R = S and go to Step 2.
   Otherwise increment j by one. If $j \leq MAX_{neigh}$ go to Step 3.
5. When $j > MAX_{neigh}$, compare the cost of S with the best solution found so far. If the cost of S is less, record this cost and the representation.

   Increment i by one. If $i > MAX_{sol}$ stop, otherwise go to Step 1.

## 3.5 BIRCH

BIRCH builds a hierarchical data structure to incrementally cluster the incoming points using the available memory and minimizing the amount of I/O required. The complexity of the algorithm is O(N) since one pass suffices to get a good clustering though, results can be improved by allowing several passes.

BIRCH uses a hierarchical data structure called Clustering Feature tree (CF tree) for partitioning the incoming data points in an incremental and dynamic way. This section is organized as follows. First we give an overview of the four phases in the BIRCH algorithm and the construction of the CF tree. Theoretical basics of Clustering Features are given and it is described how distance metrics can work solely on Clustering Features. We finish this section by revisiting how BIRCH can fit large datasets into RAM using these Clustering Features to compress the data.

BIRCH employs four difierent phases during each clustering process.
1. Linearly scan all data points and insert them in the CF tree as described earlier.
2. Condense the CF tree to a desirable size depending on the clustering algorithm employed in step three. This can involve removing outliers and further merging of clusters.
3. Employ a global clustering algorithm using the CF tree's leaves as input. The Clustering Features allow for effective distance metrics. This is feasible because the CF tree is very densely compressed at this point. Here an agglomerative hierarchial clustering algorithm is applied directly to the subclusters represented by their CF vectors. It also provides the flexibiltiy of allowing the user to specify either the desired number of clusters or the desired diameter threshold for clusters. After this step we obtain a set of clusters that captures major distribution pattern in the data. However there might exist minor and localized inaccuracies which can be handled by an optional step 4

4. Optionally refine the output of step three. All clusters are now stored in memory. If desired the actual data points can be associated with the generated clusters by reading all points from disk again.

**Clustering Feature :** Given N d-dimensional data points in a cluster, Xi, CF vector of the cluster is defined as a triple CF = (N,LS,SS), where LS is the linear sum and SS is the square sum of data points.

**CF tree :** A CF tree is a height balanced tree with two parameters: branching factor B and threshold T. Each non-leaf node contains at most B entries of the form [CFi,childi], where childi is a pointer to its ith child node and CFi is the subcluster represented by this child. A leaf node contains at most L entries each of the form [CFi] . It also has to two pointers prev and next which are used to chain all leaf nodes together. The tree size is a function of T. The larger the T is, the smaller the tree is. We also require a node to fit in a page of size of p. B and L are determined by P. So P can be varied for performance tuning. It is a very compact representation of the dataset because each entry in a leaf node is not a single data point but a sub-cluster.

## 3.6 CURE

Hierarchical methods decompose a dataset into a tree-like structure. Two common ones are HAC and CURE. Since these methods are designed to discover clusters of arbitrary shape, they do not necessarily optimize SSQ. CURE (Clustering Using REpresentatives) is an efficient data clustering algorithm for large databases that is more robust to outliers and identifies clusters having non-spherical shapes and wide variances in size.

With the partitional clustering algorithms, which uses the sum of squared errors criterion when there are large differences in sizes or geometries of different clusters, the square error method could split the large clusters to minimize the square error which is not always correct. Also, with hierarchic clustering algorithms these problems exist as none of the distance measures between clusters (dmin,dmean) tend to work with different shapes of clusters. Also the running time is high when n is very large. The problem with the BIRCH algorithm is that once the clusters are generated after step 3, it uses centroids of the clusters and assign each data point to the cluster with closest centroid. Using only the centroid to redistribute the data has problems when clusters do not have uniform sizes and shapes.

To avoid the problems with non-uniform sized or shaped clusters, CURE employs a novel hierarchical clustering algorithm that adopts a middle ground between the centroid based and all point extremes. In CURE, a constant number c of well scattered points of a cluster are chosen and they are shrunk towards the centroid of the cluster by a fraction α. The scattered points after shrinking are used as representatives of the cluster. The clusters with the closest pair of representatives are the clusters that are merged at each step of CURE's hierarchial clustering algorithm. This enables CURE to correctly identify the clusters and makes it less sensitive to outliers. The running time of the algorithm is $O(n^2 \log n)$ and space complexity is $O(n)$.

The algorithm cannot be directly applied to large databases. So for this purpose we do the following enhancements

**Random sampling:** To handle large data sets, we do random sampling and draw a sample data set. Generally the random sample fits in main memory. Also because of the random sampling there is a trade off between accuracy and efficiency.

**Partitioning for speed up**: The basic idea is to partition the sample space into p partitions. Then in the first pass partially cluster each partition until the final number of clusters reduces to np/q for some constant q ≥ 1. Then run a second clustering pass on n/q partial clusters for all the partitions. For the second pass we only store the representative points since the merge procedure only requires representative points of previous clusters before computing the new representative points for the merged cluster. The advantage of partitioning the input is that we can reduce the execution times.

**Labeling data on disk:** Since we only have representative points for k clusters, the remaining data points should also be assigned to the clusters. For this a fraction of randomly selected representative points for each of the k clusters is chosen and data point is assigned to the cluster containing the representative point closest to it.

CURE(no. of points, k)

Input : A set of points S

Output : k clusters

1. For every cluster u (each input point), in u.mean and u.rep store the mean of the points in the cluster and a set of c representative points of the cluster initially c = 1 since each cluster has one data point. Also u. closest stores the cluster closest to u.
2. All the input points are inserted into a k-d tree T.
3. Treat each input point as separate cluster, compute u.closest for each u and then insert each cluster into the heap Q.
4. While size(Q) > k.
5. Remove the top elemnt of Q(say u) and merge it with its closest cluster u.closest(say v) and compute the new representative points for the merged cluster w. Also remove u and v from T and Q.
6. Also for all the clusters x in Q, update x.closest and relocate x.
7. Insert w into Q.
8. Repeat.

Hierarchical algorithms, including BIRCH are known to suffer from the problem that hierarchical merge or split operations are irrevocable.

## 3.7 STREAM

STREAM is an algorithm for clustering data streams described by Guha, Mishra, Motwani and O'Callaghan which achieves a constant factor approximation for the k-Median problem in a single pass and using small space.

To understand STREAM, the first step is to show that clustering can take place in small space. Small-Space is a divide-and-conquer algorithm that divides the data, S, into pieces, clusters each one of them (using k-means) and then clusters the centers obtained.

**Small-Space Algorithm representation**

Algorithm Small-Space(S)

1. Divide S into *l* disjoint pieces X1,...,X.

2. For each i, find O(k) centers in Xi. Assign each point in Xi to its closest center.

3. Let X' be the O(*l*k) centers obtained in (2), where each center c is weighted by the number of points assigned to it.

4. Cluster X' to find k centers.

We can also generalize Small-Space so that it recursively calls itself i times on a successively smaller set of weighted centers and achieves a constant factor approximation to the k-median problem.

The problem with the Small-Space is that the number of subsets that we partition S into is limited, since it has to store in memory the intermediate medians in X'. So, if M is the size of memory, we need to partition S into subsets such that each subset fits in memory, (n/) and so that the weighted k centers also fit in memory, k<M. But such an may not always exist.

The STREAM algorithm solves the problem of storing intermediate medians and achieves better running time and space requirements. The algorithm works as follows:

1. Input the first m points; using the randomized algorithm presented in [3] reduce these to O(k) (say 2k) points.
2. Repeat the above till we have seen m2/(2k) of the original data points. We now have m intermediate medians.
3. Using a local search algorithm, cluster these m first-level medians into 2k second-level medians and proceed.
4. In general, maintain at most m level-i medians, and, on seeing m, generate 2k level-i+ 1 medians, with the weight of a new median as the sum of the weights of the intermediate medians assigned to it.
5. When we have seen all the original data points, we cluster all the intermediate medians into k final medians, using the primal dual algorithm.

## 3.8 LSEARCH

STREAM needs a simple, fast, constant-factor approximation k-Median subroutine. We believe ours is the first such algorithm that also guarantees flexibility in k. The LSEARCH algorithm does not directly solve k-Median but could be used as a subroutine to a k-Median algorithm, as follows. We first set an initial range for the facility cost z between 0 and an easy-to-calculate upper bound; we then perform a binary search within this range to find a value of z that gives us the desired number k of facilities; for each value of z that we try, we call Algorithm CG to get a solution.
Algorithm CG(data set N, facility cost z)
1. Obtain an initial solution (I; f) (I fi N of facilities, f an assignment function) that gives a n-approximation to facility location on N with facility cost z.
2. Repeat (log n) times:
   - Randomly order N.
   - For each x in this random order: calculate gain(x), and if gain(x) > 0, add a facility there and perform the allowed reassignments and closures.

For a given facility location instance, there may exist a k such that there is no facility cost for which an optimal solution has exactly k medians, but if the dataset is "naturally k-clusterable," then our algorithm should find k centers.
On each iteration, we expect the total solution cost to decrease by some constant fraction of the way to the best achievable cost, if our initial solution is a constant-factor approximation rather than an n-approximation as used by Charikar and Guha, we can reduce our number of iterations from $\Theta$ (log n) to $\Theta(1)$.
Initial Solution (data set N, facility cost z)
1. Reorder data points randomly
2. Create a cluster center at the first point
3. For every point after the first,
   - Let d be the distance from the current data point to the nearest existing cluster center
   - With probability d=z create a new cluster center at the current data point; otherwise add the current point to the best current cluster.

This algorithm runs in time proportional to n times the number of facilities it opens and obtains an expected approximation to optimum.
Next we need to Obtain Feasible Centers, Assume the points $c_1,....., c_k$ constitute an optimal solution to the k-Median problem for the dataset N, that $C_i$ is the set of points in N assigned to $c_i$, and that $r_i$ is the average distance from a point in $C_i$ to $c_i$ for $1 \leq i \leq k$.

Therefore, we will give a **Facility Location** subroutine that our k-Median algorithm will call; it will take a parameter $\varepsilon \in \Re$ that controls how soon it stops trying to improve its solution.

158

Algorithm FL(N, d(.,.), z, $\varepsilon$, (I, a))

1. Begin with (I,a) as the current solution

2. Let C be the cost of the current solution on N.

Consider the feasible centers in random order, and for each feasible center y, if gain(y) > 0, perform all advantageous closures and reassignments, to obtain a new solution (I`, a`) .

3. Let C0 be the cost of the new solution; if C` be the cost of the new solution;

if C` ≤ (1 - $\varepsilon$ )C, return to step 2

Now we will give our k-Median algorithm for a data set N with distance function d.

$Algorithm\ LSEARCH(N, d(.,.), k, \varepsilon, \varepsilon^1, \varepsilon^n)$

1. $Z_{min} \leftarrow 0$

2. $Z_{max} \leftarrow \sum x \in Nd(x, x_0) (for\ x_0\ an$ arbitrary point in N)

3. $Z \leftarrow (Z_{max} + Z_{min})/2$

4. $(I, a) \leftarrow InitialSolution(N, z).$

5. $Randomly\ pick\ \theta(\frac{1}{p}\log k)\ points\ as\ feasible\ medians$

6. While # medians # k and $Z_{min} < (1 - \varepsilon^{``})Z_{max}$;

- Let (F, g) be the current solution
- Run FL(N, d, $\varepsilon$, (F, g)) to obtain a new solution(F`, g`)
- If $|F|$ is about k, then (F`, g`) ← FL(N, d, $\varepsilon$`, (F`, g`))
- If $|F| > k$ then { $Z_{min} \leftarrow Z$ and $Z \leftarrow (Z_{max} + Z_{min})/2$};

$else\ if\ |F| > k\ then\ \{Z_{max} \leftarrow Z\ and\ Z \leftarrow (Z_{max} + Z_{min})/2\}$

7. To simulate a continuous space

   move each cluster center to the center - of - mass for its cluster.

8. Return our solution(F`, g`).

The initial value of zmax is chosen as a trivial upper bound on the value of z we will be trying to find. The running time of LSEARCH is O(nm + nk log k) where m is the number of facilities opened by InitialSolution. m depends on the properties of the dataset but is usually small, so this running time is a significant improvement over previous algorithms.

## 4. CONCLUSION

Most of the algorithms generally assume some implicit structure in the data set. The problem, however, is that usually you have little or no information regarding the structure, which is, paradoxically, what you want to uncover. Another issue to keep in mind is the kind of input and tools that the algorithm requires. An additional issue related to selecting an algorithm is correctly choosing the initial set of clusters. This paper provides an broad survey of the most basic techniques, and as is stated on the title, an overview of the elementary clustering techniques most commonly used.

 Author

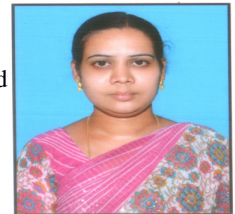

**T. Soni Madhulatha** obtained MCA from Kakatiya University in 2003 and M. Tech (CSE) from JNTUH in 2010.She has 8 years of teaching experience and she is presently working as Associate professor in  Department of Informatics in Alluri Institute of Management Sciences, Hunter Road Warangal. She published papers in various National and  International Journals and Conferences. She is a Life Member of ISTE ,  IAENG and APSMS.